# Analyzing and Calibrating Risk Assessment by Software Developers

Yukasa Murakami
*Kindai University*
Osaka, Japan

Masateru Tsunoda
*Kindai University*
Osaka, Japan
tsunoda@info.kindai.ac.jp

Eduardo C. Campos
*Federal University of Uberlândia*
Uberlândia, Brazil
eduardocunha11@gmail.com

*Abstract*—In software project management, risk management is a critical factor. Project managers use existing lists of risk or perform brainstorming to identify the risks. However, it is not easy to perceive all the risks objectively. As a result, some risks are perceived based on subjective impression, which leads to risk biases. So, our goals are (i) We clarify the risk perception of developers to enhance the reliability of the brainstorming, and (ii) we calibrate the risk assessment based on a mathematical model to make more accurate risk list. In the analysis, we collected data concerning the risk perception of 69 professional software developers via a questionnaire. The average number of years of experience among these professionals was 18.3. Using the dataset, we applied factor analysis to clarify the factors that affect the evaluation of risk impact. The questionnaire was based on the risk perception theory established by Slovic, in which "dread" and "unknown" are the major factor of risk perception. The analysis result shows that (i) risk experience (i.e., whether a developer actually faced the risk or not) sometimes affects risk assessment (evaluation of risk impact), (ii) risk perception is considered to be based on unknown and dread factors, and (iii) risk assessment can be calibrated by a mathematical model (the average absolute error was 0.20).

## I. INTRODUCTION

*How do we ensure that self-driving cars, nuclear power plants, railway systems and heart monitors are safe and reliable?* That is the topic of risk management. Fault tree analysis [25] is an exceedingly popular technique in this context, employed by many institutions such as NASA, ESA, Honeywell, Ford, Airbus, Toyota, and Shell.

In software project management, risk management is one of the important factors. For example, project risk management is one of the knowledge areas of the Project Management Body of Knowledge (PMBOK) [13]. According to the Project Management Institute's PMBOK, risk management is one of the ten knowledge areas in which a project manager must be competent[1]. Risk management procedures consist of identifying, prioritizing, planning, monitoring and controlling the risks. Project managers use existing risk lists or perform brainstorming to identify the risks. Some studies such as [26] involved asking project managers about the risks of software development projects and their impact, in order to develop these risk lists.

Generally, it is difficult to perceive all the risks objectively. As a result, some risks are perceived based on subjective impression, which could lead to risk biases. Risk biases refer to the difference between the objective risk (i.e., probability of the risk) and the subjective risk [12]. For example, the probability of the accidents of aircraft is far lower than that for automobile accidents. However, several people do not recognize this difference owing to the risk biases.

In this study, we assume that the perception of the risks is affected by actual experience with the risks. As a result, risk identification may be biased by the experience. The risk lists are often made by a survey, and it includes major risks (e.g., insufficient/inappropriate staffing) to help risk assessment. For example, when the risk list is made but subjects do not have enough experience of some risks, the list will have some biases (i.e., some major risks may be overlooked). Similarly, when the brainstorming is conducted but project members do not have enough experience of some risks, the result will be also biased.

To cope with the bias on the identification of the risks, we focus on a principal study regarding risk perception conducted by Slovic [33]. He identified three factors that affected the perception of 81 risks. A series of studies on risk perception clarified that "dread risk" and "unknown risk" are the two major factors concerning risk perception, and these factors affect the risk biases. We will decrease the risk biases on the brainstorming of risk identification explained above, when one should be explicitly aware of these two factors, and the accuracy and correctness of the risk perception and risk assessment based on these factors should be discussed. For instance, if a project member assigns some weight to "dread" when evaluating the impact of a risk, the project members should discuss the correctness of this perception.

Also, it is expected to calibrate the risk list based on answers of subjects when subjects do not have enough experience of some risks. The list is calibrated by the mathematical model which uses answers of the dread and unknown factors as independent variables. The goal of our study is to calibrate the risk assessment based on the analysis of the risk perception of the developers. So, we investigated software project risks via a questionnaire, and analyzed the risk perception of software developers. The main contributions of the study are to clarify the following aspects by analyzing the responses of many professional developers:

- Risk experience sometimes affects risk assessment (evaluation of risk impact) (in subsection III.A);
- Risk perception is considered to be based on unknown and dread factors (in subsection III.C);
- Risk assessment can be calibrated by mathematical models (in section IV)

---

[1] Practice Standard Project Risk Management". PMI. http://www.pmi.org

Although we analyzed only two risks, they are two out of 11 major risks shown by Keil et al. [17], and therefore the analysis results is expected to be applicable to some risks. The data is precious because our data was collected from many professional developers, although many studies do not use such data (i.e., the data was collected from students, and the number of professional subjects was smaller than our study on many studies). Sjoberg et al. [32] surveyed experiments in software engineering studies, and indicated that 72.6 percent of the experiments adopted only students as subjects, 18.6 percent adopted only professionals, and 8.0 percent adopted mixed group. Mean of the subjects of them were 56.0, 20.0, and 49.3 respectively.

It seems natural that the risk assessment is affected by actual experience with the risks (i.e., whether a developer actually faced the risk or not). However, past studies that investigated software project risks using a questionnaire (e.g., [19]) did not consider the risk experience, confirming that the experience was not regarded indispensable in past studies. In empirical software engineering, it is important to confirm the developers' knowledge by quantitative data analysis [15]. This is because such knowledge is sometimes regarded as incorrect, based on the quantitative data analysis.

The established theories in other fields are not always applicable to the field of software engineering. Therefore, some studies analyzed the applicability of such established theories to the software engineering field [16]. From the viewpoint, this paper may be regarded as a replicated study, and major contribution of the replication is that it was performed asking large numbers of professional software engineers. In addition, we analyzed the feasibility of calibration of risk assessment (see section IV).

## II. DATA COLLECTION

### A. Investigation by Questionnaire

To analyze the risk perception of software engineers, we applied the questionnaire method [8], which is often used in psychology. In the questionnaire, the developers studied two risks, and responded to the questions asked in the questionnaire by stating the extent to which they agree or disagree with each presented statement. **Q3 … Q11** are five grade scales (1: agree …5: disagree). The details of each of these questions are as follows:

1) **Risk A**: Failure to gain user commitment;
2) **Risk B**: Insfficient/inappropriate staffing;

- **Question 1 (Q1)**: Impact to success / failure of software project: 10 grade (1: less impact . . . 10: more impact);

TABLE I. NUMBER OF EMPLOYEES OF RESPONDENTS' ORGANIZATION

|  | Number of respondents | Ratio |
| --- | --- | --- |
| Under 300 | 26 | 38% |
| 300 … 1,000 | 10 | 14% |
| Over 1,000 | 33 | 48% |

TABLE II. AVERAGE TEAM SIZE OF A PROJECT IN THE ORGANIZATION OF THE RESPONDENTS

|  | Number of respondents | Ratio |
| --- | --- | --- |
| Under 10 | 27 | 39% |
| 10 … 50 | 22 | 32% |
| Over 50 | 20 | 29% |

- **Question 2 (Q2)**: Frequency of your own experience of risks: 3 grades (1: rarely, 2: occasionally, 3: often)
- **Question 3 (Q3)**: The risk is difficult to control by developers;
- **Question 4 (Q4)**: Damage to stakeholders is large when the risk occurs;
- **Question 5 (Q5)**: Several stakeholders suffer when the risk occurs;
- **Question 6 (Q6)**: Probability of the risk is increasing;
- **Question 7 (Q7)**: The occurrence of risk causes your own dismissal or bankruptcy of your company;
- **Question 8 (Q8)**: The influence is observed immediately when the risk occurs;
- **Question 9 (Q9)**: Stakeholders have accurate knowledge regarding the risk;
- **Question 10 (Q10)**: The influence of the risk to the project is clarified scientifically;
- **Question 11 (Q11)**: The risk is a new problem that has manifested in recent years

**Risk A** and **Risk B** were selected from the major risks to software projects, listed by Keil et al. [19]. We focused on high and low ranked risks in the list, because we assumed that the rank is affected by the risk perception of developers, and therefore, the difference in the perception is large between high and low rank risks. When the difference is large, the bias can be analyzed more easily. Note that low rank risks are still one of the major risks. Next, from the risks, we selected the risks that were easy to understand and for which the risk recognition of the developers was considered to be different (i.e., controllability by developers was different).

Q3 … Q7 are questions related to the "dread risk". Q8 … Q11 are questions related to the "unknown risk". In Slovic's study [33], the factor "dread risk" related to 10 risk characteristics (e.g., controllable, global catastrophic, and high risk to future generations). The factor "unknown risk" related to five risk characteristics (e.g., new risk, and risk to known science). To enhance the response rate and reliability of the response, we disregarded the characteristics that did not relate directly to software development, and the risks that are difficult to explain briefly (e.g., high risk to future generations). It should be noted that studies on risk perception have been widely conducted, and recent studies [1, 11] did not consider all the risk characteristics explained above. Similarly, we did not consider all the characteristics.

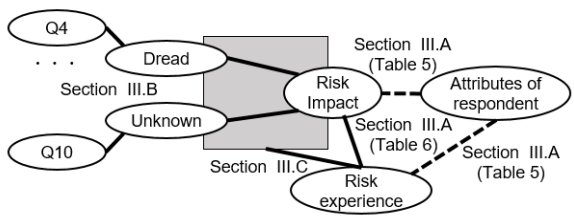

Fig. 1. Overview of the analysis

TABLE III. PROCESS TYPES IN WHICH RESPONDENTS WERE ENGAGED

| Process Type | Number of respondents |
| --- | --- |
| Development | 19 |
| Testing | 17 |
| Project management | 13 |
| Development support | 14 |
| Research | 10 |

### B. Summary of Respondents' Attributes

We collected data from the participants of Software Symposium (SS)[2] which was held in Japan, 2017. Most of the participants were software developers employed by various companies. The number of respondents was 69.

Table 1 … 3 lists the basic statistics of attributes of the respondents. The average years of experience was 18.3, and the median was 16.0 years. The developers had sufficient experience, based on the average years of experience. The distribution of the size of organizations that employed these developers is somewhat biased in the Table 2 (The number of small and large organizations was large, compared with that of medium organizations). However, the distribution of project size was not biased. Table 3 lists the distribution of development process in which respondents were engaged.

As explained in subsection III.A, the relationship between respondents' attributes such as the size of respondents' organization and the evaluation of impact of risks (Q1) was weak. Therefore, we did not remove any data points based on the attributes.

### III. ANALYZING RISK PERCEPTION

Figure 1 shows an overview of the analysis. In subsection III.A, we analyzed the influence of the attributes of the respondents. Then, we analyzed the relationship between risk evaluation (Q1) and risk evaluation (Q2). In subsection III.B, using factor analysis (Table 7), we clarified risk perception (i.e., dread risk and unknown risk). In subsection III.C, we analyzed the influence of the experience of risks (Q2) on the relationship between risk perception (dread risk and unknown risk) and risk evaluation (Q1).

### A. Relationship between Experience and Evaluation of Risk

**Influence of respondents' attributes**: Risk evaluation (i.e., answer to Q1) may be affected by the attributes of respondents such as years of development experience. For example, a developer who has 10 years of experience answers 7 to Q1 of risk A. However, a developer who has 15 years of experience answers 10. In this example shown in Table 4, the years of experience affects the answer to Q1 (i.e., there is a positive relationship between the years of experience and Q1). To clarify such a relationship, we analyzed the relationship between the evaluation of risk and attributes of respondents, using the Pearson correlation coefficient and correlation ratio. The correlation ratio is used to analyze the relationship between a nominal variable (i.e., the attributes listed in Table 3) and a numerical variable. The correlation coefficient was calculated by pairing the years of development experience with Q1 (see Table 5). The correlation ratio was calculated between development support (whether respondents were engaged in development support or not; a dummy variable) with Q1.

As listed in Table 5, the strength of each relationship was small (basic statistics of each attribute are shown in Tables 1, 2, and 3). Therefore, risk evaluation (Q1) and risk experience (Q2) are not considered to be affected by the attributes of developers such as years of experiences. Therefore, we do not consider the influence of respondents' attributes when analyzing risk evaluation and risk experience.

**Influence of experience of risk**: When a developer actually faces risk, the evaluation of the influence of risk on software project may change. Thus, we analyzed the relationship between risk evaluation (answer to Q1) and frequency of risk experience (answer to Q2). To analyze the relationship, we stratified the dataset based on the frequency of experience and calculated the average risk evaluation for each risk (i.e., risks A and B). We did not use the correlation coefficient because the relationship between risk evaluation and risk experience was not linear, and the relationship was different for each risk. This may be affected by interpretation of "2: occasionally", one of the choices included in Q2 (frequency of experience).

Table 6 summarizes the results of the analysis. For risk A, developers evaluated that the influence of risk was small when the frequency of experience was small. The differences between the average risk evaluation was approximately 1.0 for each frequency of experience (i.e., rarely, occasionally, and often), and the difference was statistically significant (p-value was 0.00). In contrast, for **risk B**, when the frequency of experience is low, the evaluation of the influence of risk was slightly higher (Note that the difference was not statistically significant (p-value was 0.72)). That is, for a high-ranked risk in the list [17], when the frequency of experience was low, the evaluation of the influence of risk was also low. For a low-ranked risk in the list, the tendency was opposite. The result suggests that the frequency of experience is considered to be sometimes related to the evaluation of the influence of risk.

---

[2] Software Symposium (SS' 2017). http://sea.jp/ss2017

TABLE IV. AN EXAMPLE OF IMPACT OF RISK AND ATTRIBUTES ON EACH RESPONDENT

| Respondent | Q1 (Impact of risk) | Years of experience | Development support |
|---|---|---|---|
| R1 | 7 | 10 | Yes |
| R2 | 10 | 15 | No |
| … | … | … | … |

TABLE V. RELATIONSHIPS BETWEEN RESPONDENTS' ATTRIBUTES AND IMPACT OF THE RISKS OR FREQUENCY OF RISK EXPERIENCE

| Attribute | Impact of the Risks | Frequency of Risk Experience |
|---|---|---|
| Number of employees | 0.16 | -0.13 |
| Years of experience | 0.16 | 0.04 |
| Average team size | 0.08 | 0.01 |
| Development | 0.02 | 0.06 |
| Testing | 0.06 | 0.02 |
| Project management | 0.13 | 0.06 |
| Development support | 0.04 | 0.10 |
| Research | 0.08 | 0.00 |

TABLE VI. EVALUATION OF RISK IMPACT STRATIFIED BY FREQUENCY OF RISK EXPERIENCE

| Risk A/B | Frequency of experience | Average | Number of respondents | Standard deviation |
|---|---|---|---|---|
| A | 1 (Rarely) | 6.4 | 22 | 3.0 |
|   | 2 (Occasionally) | 7.7 | 29 | 2.0 |
|   | 3 (Often) | 8.6 | 18 | 1.4 |
|   | Total | 7.5 | 69 | 2.4 |
| B | 1 (Rarely) | 8.4 | 5 | 1.7 |
|   | 2 (Occasionally) | 8.3 | 35 | 1.7 |
|   | 3 (Often) | 8.0 | 29 | 2.2 |
|   | Total | 8.2 | 69 | 1.9 |

TABLE VII. FACTOR PATTERN AFTER PROMAX ROTATION

| Questions | I | II |
|---|---|---|
| Q04 | 0.84 | -0.03 |
| Q05 | 0.73 | -0.01 |
| Q08 | 0.31 | 0.26 |
| Q07 | 0.29 | 0.19 |
| Q03 | 0.15 | 0.11 |
| Q06 | 0.14 | 0.02 |
| Q11 | 0.09 | 0.00 |
| Q10 | -0.13 | 0.77 |
| Q09 | 0.19 | 0.57 |

*B. Factors of Risk Perception*

We applied factor analysis to derive major factors that summarize risk perception. The analysis procedure was similar to that generally applied to make psychological scale in psychology [21]. First, to make the scale, questions to measure constructs (i.e., dread and unknown) are made (i.e., Q3 … Q11). Then, using factor analysis, the questions which do not relate to the constructs very much are eliminated, as shown in Figure 1. Finally, the reliability of the scale is evaluated by Cronbach's alpha.

**Applicability of factor analysis**: First, we checked applicability of factor analysis to the dataset collected in this study by using Kaiser–Meyer–Olkin (KMO) and Bartlett's test of sphericity. They are used to clarify whether there is any relationship between the variables. When KMO is larger than 0.5 and Bartlett's test of sphericity is significant, there are relationships between variables [24]. Therefore, we can apply factor analysis. On the dataset, KMO was 0.56 and the p-value of Bartlett's test of sphericity was 0.00. Thus, factor analysis can be applied to the dataset.

**Decision of the number of factors**: Next, we decided the number of factors. The number of factors is decided by selecting factors whose eigenvalues were not smaller than 1 [10], or applying preset value that has already been established by past studies. When we selected factors whose eigenvalues were not smaller than 1, the number of factors was four. However, communality was larger than 1. This means that the analysis is not proper. When we set the number of factors as three, only Q6 of factor loading was larger than 0.4 (standard criterion [14]) on the third factor. Additionally, the relationship between the third factor and the evaluation of the influence of risk was weak. Therefore, we set the number of factors as two in the following analysis.

Note that although in Section II, we assumed two factors (i.e., dread risk and unknown risk) to settle the questions Q3 … Q11, it is not natural that two factors were identified by the factor analysis. In psychology, factors are considered first, and questions are generated based on the factors [30], similar to the procedure in our study. However, if the questions are not appropriate, factor analysis cannot be applied (i.e., the results of KMO and Bartlett's test of sphericity indicates inappropriateness of factor analysis).

TABLE VIII. RELATIONSHIPS BETWEEN EVALUATION OF RISK IMPACT AND SUBSCALE SCORE

| Frequency of experience | | 3 (Often) | 2 (Occasionally) | 1 (Rarely) |
|---|---|---|---|---|
| Dread factor | Correlation coefficient | 0.59 | 0.28 | 0.16 |
| | p-value | 0.00 | 0.03 | 0.42 |
| | Number of respondents | 47 | 64 | 27 |
| Unknown factor | Correlation coefficient | 0.45 | 0.10 | 0.28 |
| | p-value | 0.00 | 0.45 | 0.16 |
| | Number of respondents | 47 | 64 | 27 |

TABLE IX. AVERAGE OF SUBSCALE SCORES STRATIFIED BY FREQUENCY OF EXPERIENCE

| Risk A/B | Frequency of experience | | Impact of risk | Dread factor | Unknown factor |
|---|---|---|---|---|---|
| A | 1 (Rarely) | Average | 6.4 | 4.1 | 2.5 |
| | | Number of respondents | 22 | 22 | 22 |
| | | Standard deviation | 3.00 | 1.08 | 0.69 |
| | 2 (Occasionally) | Average | 7.7 | 4.4 | 2.4 |
| | | Number of respondents | 29 | 29 | 29 |
| | | Standard deviation | 1.97 | 0.60 | 0.83 |
| | 3 (Often) | Average | 8.6 | 4.3 | 2.7 |
| | | Number of respondents | 18 | 18 | 18 |
| | | Standard deviation | 1.38 | 0.91 | 1.09 |
| B | 1 (Rarely) | Average | 8.4 | 4.6 | 3.2 |
| | | Number of respondents | 5 | 5 | 5 |
| | | Standard deviation | 1.67 | 0.55 | 0.84 |
| | 2 (Occasionally) | Average | 8.3 | 4.1 | 2.8 |
| | | Number of respondents | 35 | 35 | 35 |
| | | Standard deviation | 1.74 | 0.80 | 0.93 |
| | 3 (Often) | Average | 8.0 | 3.9 | 2.8 |
| | | Number of respondents | 29 | 29 | 29 |
| | | Standard deviation | 2.16 | 1.14 | 0.99 |

TABLE X. SUMMARY OF PREDICTION MODELS OF RISK IMPACT

| Model | Dependent variable | Method | Independent variables | Input values |
|---|---|---|---|---|
| $R1_{A,B}$ | Q1 | Regression | Q2 | 3 |
| $M1_{A,B}$ | Q1 | Mean | - | - |
| $R2_{A,B}$ | Dread | Regression | Q2 | 3 |
| $M2_{A,B}$ | Dread | Mean | - | - |
| $R3_{A,B}$ | Unknown | Regression | Q2 | 3 |
| $M3_{A,B}$ | Unknown | Mean | - | - |
| R4R | Q1 | Regression | Q2, dread, unknown | 3, $R2_{A,B}$, $R3_{A,B}$ |
| R4M | Q1 | Regression | Q2, dread, unknown | 3, $M2_{A,B}$, $M3_{A,B}$ |

**Selecting and naming factors**: Lastly, we selected two factors and named them. We used unweighted least-squares as the factor extraction method (It is used to build a model). We applied promax rotation (It is used to interpret the built model).

The ratio of variance explained of all questions by the two factors was 40.7%. The correlation coefficient between the factors was 0.19, and the value was relatively small. The results suggest that the performance of the factor analysis was appropriate to some extent. Table 7 lists the factor pattern after rotation. Italic numbers denote factor loadings greater than 0.4. The factor loadings of Q4 and Q5 were high for the first factor identified in the above analysis. The factor loadings of Q9 and Q10 were high for the second factor. Q4 and Q5 are questions on risk characteristics (see section II) related to dread risk identified in Slovic's study [33]. Q9 and Q10 are questions on the characteristics relating to unknown risk. Hence, we named the first factor as "dread factor" and the second factor as "unknown factor".

To evaluate the reliably of the dread factor and the unknown factor, we used Cronbach's alpha. When the alpha *gt* 0.5, it is regarded as acceptable [9]. The alpha of the dread factor was 0.77 and that of the unknown factor was 0.58. So, the factors are reliable.

*C. Relationship between Risk Experience and Perception*

We analyzed the relationship between the factors (average answers to questions selected by the factor analysis) and the frequency of experience. First, we calculated the subscale score to summarize the answers to 9 questions (Q3 … Q11) to "dread" and "unknown" factors for each respondent. The calculation is based on the results of the factor analysis. To calculate the score, questions whose factor loadings are high are selected, and the average or sum of answers to the questions is used. In the analysis, we used the average of answers to Q4 and Q5 (see subsection III.B) as the subscale score of the dread factor. Furthermore, we used the average of answers to Q9 and Q10 as the subscale score of the unknown factor.

In the next step, we focused on experienced and non-experienced developers separately and analyzed to which factor they gave weight, when risk impact was evaluated. To do so, we stratified the dataset based on the frequency of experience and compared the correlation coefficients (i.e., strength of relationships) of risk impact (Q1) and each factor (subscale scores of the dread and unknown factors). Although there is no strict criterion for the correlation coefficient, when the absolute value of the correlation coefficient varies from 0.2 to 0.4, the relationship is often regarded as weak in health care sciences [35] (In social sciences, when the value varies from 0.15 to 0.3, it is regarded as a weak relationship, while a value varying from 0.4 to 0.6, it is regarded as weak in natural sciences [34]).

Table 8 shows the correlation coefficient stratified by the frequency of experience. It denotes relationship between the risk impact (i.e., Q1) and the factors. When the frequency was "3 (often)", correlation coefficients of both factors (subscale scores) were statistically significant, and the correlation coefficient of dread was larger. When the frequency was "2 (occasionally)", only the correlation coefficient of dread factor was statistically significant, and the correlation coefficient of unknown was very small. When the frequency was "1 (rarely)", both correlation coefficients were not statistically significant. This may be because the number of data points was small. The correlation coefficient of unknown was larger than that of dread in this case.

Dread factor (subscale score) is the average of extent of damage (Q4) and range of damage (Q5). Therefore, experienced respondents are considered to evaluate the impact of risk, giving weight to the damage to project and stakeholders. In contrast, non-experienced respondents are regarded to give weight to unknown aspects of risk, when evaluating the impact. Since unknown factor was average of knowledge of risk (Q9 and Q10).

Lastly, we further investigated on how non-experienced and experienced respondents evaluated risk A and risk B, based on the analysis (i.e., experienced developers focused on the dread factor of risk, and non-experienced developers focused on the unknown factor of risk). As shown in Table 9, we calculated the average of subscale scores of the risks, stratifying the dataset according to the kind of risks. Non-experienced respondents evaluated that the impact of risk A (user commitment) is higher than that of risk B (inappropriate staffing). This is because the unknown factor of risk B was larger than that of risk A, although the dread factor of risk A was larger than that of risk B. That is, non-experienced respondents are considered to give weight to unknown factor, when evaluating impact of the risks. For experienced respondents, the dread factor of risk A was slightly higher than that of risk B. This would affect the fact that the impact of risk A was slightly higher than that of risk B. Note that we did not statistically test the difference, since we mainly focus on the correlation coefficients shown in Table 8, and we refer Table 9 to discuss the results.

The dataset used in the analysis was collected from expert developers (see subsection II.B). However, the evaluation of the impact of risks (especially risk A) was different between developers for whom the frequency of experience was often (i.e., Q2 = 3), and developers for whom the frequency was rarely (i.e., Q2 = 1). Slovic's study [33] illustrated the difference between experts and non-experts. In contrast, the result of our study suggests that risk perception of expert software developers may be biased, if they do not actually face the risk.

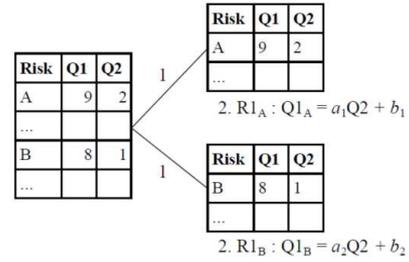

Fig. 2. Procedure to Build Model $R1_A$ and $R1_B$

IV. CALIBRATING RISK ASSESSMENT BY EXTRAPOLATION

A. Overview

When making the risk list based on a questionnaire (e.g., [19]), there may be no or few developers whose frequency of experience of risk was often (Q2 = 3). So, we evaluated the feasibility of predicting risk impact (Q1) evaluated by them. Certainly, developers whose frequency was often may be also biased. However, we can only use the evaluation of developers to assess risk, and their evaluation is considered to be reliable comparatively. Similarly, some studies proposes methods to calibrate survey results with statistical models [20].

The easiest way to predict the risk impact is using the mean of risk impact evaluated by developers whose frequency of experience of risk was rarely or occasionally (i.e., Q2 = 1 or 2). Also, we can predict the evaluation by extrapolation with linear regression. That is, Q2 is used as an independent variable, and linear regression clarified the linearity between Q1 and Q2 when Q2 = 1 and 2. Next, '3 (often)' is input to Q2 to predict Q1. To clarify the purpose of the analysis, we set the following research questions:

- **RQ1**: Which method (i.e., linear regression or mean) is better to predict the risk evaluation?
- **RQ2**: For software projects with a few highly experienced developers, what value (i.e., mean of the developers or prediction) should be used to measure the success or failure of a software project?

B. Procedure of the Experiment

To clarify **RQ1**, we constructed 8 models, as shown in Table 10. Before making the models, we removed data points on which Q2 was 3. The procedure of constructing model $R1_A$ and $R1_B$ is as follows (see figure 2):

1) Removed data points on which Q2 was 3;
2) Stratified the dataset into risk A and B;
3) Built $R1_A$ and $R1_B$ model on the subset;
4) Predicted $Q1_A$ and $Q1_B$, setting Q2 as 3 (extrapolation).

Model $R2_A$, $R2_B$, $R3_A$, and $R3_B$ were made in a similar way. The models were made to predict the dread and unknown factors (subscale score). The values predicted by them were used in model R4R and R4M.

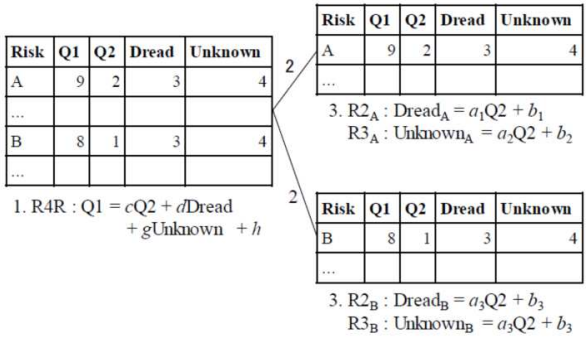

Fig. 3. Procedure to Build Model R4R

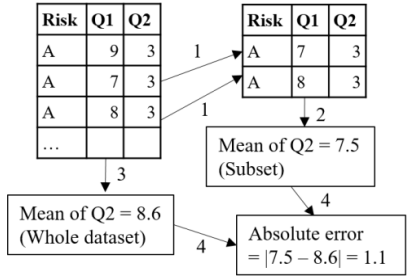

Fig. 4. Procedure to Calculate Mean of Q2 and Absolute Error

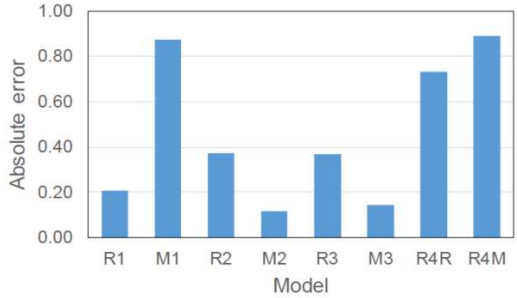

Fig. 5. Absolute Error of Each Model

Model R4R and R4M are same (i.e., their regression coefficients are same), but the values input to independent variables are different. The procedure of constructing these models is as follows (see figure 3):

1) Removed data points on which Q2 was 3;
2) Built R4R model;
3) Stratified the dataset into risk A and B;
4) Built $R2_A$, $R3_A$, $R2_B$, and $R3_B$ models on the subsets;
5) Predicted $Q1_A$, setting Q2 as 3 and using the value predicted by $R2_A$ and $R3_A$;
6) Predicted $Q1_B$, setting Q2 as 3 and using the value redicted by $R2_B$ and $R3_B$.

Model R1, R2, R3, and R4R use linear regression, and model M1, M2, M3 use arithmetic mean, to answer **RQ1**. R4M uses both.

To answer **RQ1**, we changed the number of data points where Q2 was 3, to calculate the mean of Q1 of them. The

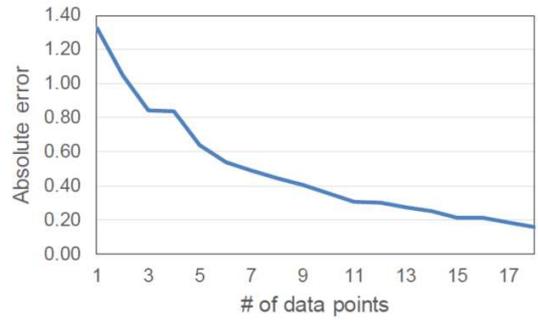

Fig. 6. Relationship between Absolute Error and Number of Data Points

procedure of constructing these models is as follows (see figure 4):

1) Randomly selected the data points where Q2 was 3,
2) Mean of Q2 on the subset was calculated,
3) Mean of Q2 on the whole dataset was calculated,
4) Absolute error between (2) and (3) was calculated,
5) Repeated (1) … (4) 20 times, and calculated average of the absolute error.

**Accuracy evaluation**: To evaluate the accuracy of the models, we used absolute error. When the difference is large, the accuracy is regarded as low. We calculated the difference between the values predicted by the models, and the mean of Q1 evaluated by developers whose frequency of experience of risk was often (Q2 = 3). That is, data fitted to build the models is the subset of the dataset where Q2 is 1 and 2. Test data to evaluate the model is the subset where Q2 is 3.

*C. Results of the Experiment*

**Summary of RQ1**: The result is shown in figure 5. In the figure, R1 (average of absolute error of $R1_A$ and $R1_B$) was the smallest among R1, M1, R4R, and R4M models (i.e., models whose dependent variable was Q1). M4R was the second lowest among the models. Although the absolute error of M2 and M3 was small, the error of R4M was larger than that of R4R. So, the answer of **RQ1** is "Linear regression is better than mean to predict risk evaluation. Especially, model R1 is the best to predict that".

**Summary of RQ2**: The result is shown in figure 6. When the number of data points in which Q2 was 3 was smaller than 5, the absolute error of model R4R was smaller than the mean of the subset. Moreover, when the number of datasets is smaller than 16, the error of model M1 was smaller than the mean of the subset. So, the answer of **RQ2** is "For software projects with a few highly experienced developers, it is better to use the predicted value as risk impact".

V. DISCUSSION

*A. Validity of the Questionnaire*

We investigated the risk perception of software developers by using a questionnaire which includes 9 items based on Slovic's study [33]. Although Slovic's work included more items, past studies did not always use all the items [1, 11] (i.e., they used around 10 items), because it is sufficient to analyze risk perception using a lesser number of items. In section III.B, using the dataset collected by the questionnaire, we applied factor analysis to reduce 9 items into two factors (dread risk and unknown risk). Next, we analyzed the relationship between the factors (calculated by weighted sum of the items) and frequency of experience of risks in subsection III.C.

Note that we used the 9 items which were scientifically established in the other fields, as a matter of practical convenience. That is, we do not assert that anything but these 9 items are not needed to understand various aspects of risk assessment of software developers. Also, we showed that the relationship between the frequency of risk experience and evaluation of risk impact is understood reasonably, based on the two factors established in the other fields (made by factor analysis). This does not mean that only these two factors affect risk assessment of software developers (i.e., there may be other factors which affect the risk assessment).

*B. Threats to Validity*

**Internal validity**: In the analysis, we clarified that the frequency of experience of risk affects the risk perception and evaluation of risk impact. The frequency of risk experience may be affected by the years of experience and development process in which the developers are engaged. That is, they may affect the frequency of experience, and the relationship affects the analysis results. However, as shown in Table 5, the relationship was very weak. Therefore, the years of experience and development process do not affect the results.

The evaluation of risk impact may be affected by an average team size. For example, if the team size is large, the risk impact (e.g., loss of cost) may also be large. However, it did not affect the risk evaluation, as shown in Table 5. Therefore, the analysis of risk impact is not affected by the average team size.

**External validity**: In the analysis, we used only two kinds of risk. We clarified that the frequency of experience affects the evaluation of risk impact. It is not clear that the frequency of experience affects any kind of risk, e.g., the frequency of experience may not affect certain risks. However, the frequency of experience affects one (i.e., risk A) out of 11 major risks shown by Keil et al. [17] at least. Therefore, we cannot ignore the influence of the frequency. The data is collected from various professional developers, and therefore, the influence will be observed on various software projects. Similarly, we analyzed risk perception of developers using only two risks. However, the theory of the risk perception is established in other fields, and we acquired similar results on two out of 11 major risks. So, we consider that software developers often perceive risks based on dread and unknown factors. The risk perception may not be always different when the frequency is different. However, for some risks, the risk perception was different, as shown in this study. Therefore, it is important to clarify the risk perception using the questionnaire, and check the difference in the risk assessment of the software projects. It will assist in smooth discussion.

Moreover, the calibration shown in section 4 was effective on two out of 11 major risks, even when the difference of assessment caused by experience was not significant (i.e., risk B). The accuracy of the prediction of the risk evaluation may not be always high at any risk, although the accuracy of the model will be high at some risks. We only show the feasibility of the prediction.

**Construct validity**: Dread factor and unknown factor are widely used in other fields, and our questionnaire is based on the past studies. Dread factor and unknown factor are very simply derived (i.e., sum of part of answers; See subsection 3.2). Therefore, the dread factor and unknown factor are considered to be properly constructed. In addition, the evaluation of risk impact (i.e., 10 grades) is the same as a past study [17]. Therefore, we think that the evaluation method is appropriate.

*C. Calibration Based on the Analysis Results*

The analysis results can be applied to the following situations:

**Brainstorming among developers**: When the project members are identifying the project risks (see section 1), the members can evaluate risk impact more precisely by the result. At first, answers to the questionnaire used in the analysis are collected from the members. After that, they refer to the summery of the answers of the questionnaire, and reconsider the risk impact based on the result of the analysis results. For example, if there is a risk for which the frequency of experience is low and the unknown factor is high, reconsideration of the impact of the risk is needed because the members may give too much weight to unknown factor.

**Investigation of risk impact**: When we collect the evaluation of risk impact from the developers to identify major risks of the software project, like Keil et al. [17], we can calibrate the risk impact mathematically using the questionnaire used in the study. To do that, the frequency of experience and questions about risk perception are collected when the risk impact is investigated. After the data collection, the frequency and questions are used as independent variables, and the risk impact is used as a dependent variable, to make a regression model. The model is expected to calibrate the risk impact. The feasibility of the calibration is evaluated above.

## VI. RELATED WORK

**Risk Management**: The majority of risk management studies deal with normative techniques of managing risk (e.g., [3], [4], [5], [6], [23]). A few studies have classified software risk items ([2], [27], [28], [31]). These studies consider software risks along several dimensions and have provided some empirically determined insights of typical software risks and their variations. For example, Keil et al. [17] asked 41 project managers in United States, Finland, and Hong Kong to clarify risk factors. As risk factors, Ropponen et al. [29] clarified six risk groups such as scheduling and management of requirements. However, these studies did not analyze risk perception based on "dread" and "unknown" factors, which Slovic clarified.

**Risk Perception**: The concept that different stakeholders perceive software projects in different ways is also well established in the literature [22]. Keil et al. [18] have demonstrated, for example, that users and project managers differ in terms of their project risk perceptions. In addition, though it is fair to assume that outsiders would see more risks than insiders, we found no empirical studies in the literature on the influence of outsiders on risk perception and behavior.

Theoretically, the purpose of a checklist is to provide a comprehensive risk profile and bring attention to risks that might otherwise be neglected [31]. From this perspective, a checklist is expected to help practitioners identify more risk factors than they would otherwise be able to identify without the aid of such a tool. Prior work has demonstrated that such tools can influence practitioners' risk perception (e.g., [7], [19], [22]). For instance, Keil et al. [19] analyzed the influence of probability of the risk when deciding to continue a software development project. In the analysis, subjects were students. The study did not use "dread" and "unknown" factors from Slovic's study, but gave subjects to probability of the risks directly. That is, the viewpoint of the analysis is different from our study.

## VII. Conclusion

It is important to manage risks in software development. To manage the risks, risks that may occur in the project are identified by some methods such as brainstorming, and risk treatment plans are made based on that. However, risk recognition is considered to include biases, which may affect the accuracy of identification of the risks. To recognize the risks precisely, we clarified the risk recognition of software developers. Our study identified the risk recognition quantitatively using the questionnaires based on Slovic's study. The procedure of the analysis was as follows:

1) We collected the answers to a questionnaire from 69 professional software developers (section II). The data is valuable in terms of its rarity (section I).
2) We clarified that the respondents' attributes such as years of experience did not affect other variables, and evaluation of risk impact was affected by the frequency of risk experience (subsection III.A).
3) Using factor analysis, the responses of the questionnaires were compressed to two factors ("dread" and "unknown") (subsection III.B).
4) We clarified that the strength of "dread" (average of Q4 and Q5) and "unknown" (average of Q9 and Q10) factors were different among risks when the experience of risk was different among the developers. The difference is considered to affect the recognition to severity of each risk (subsection III.C).
5) We evaluated the prediction accuracy of evaluation of risk impact, assuming that there are not many developers whose frequency of experience was often (section IV).

The following is the summary of the analysis results:

- Risk experience sometimes affects evaluation of risk impact. Therefore, the experience should be asked to assess the risks.

- Risk perception is considered to be based on unknown and dread factors. Risk perception tends to be different when risk experience is different. So, considering the perception, risk assessment among project members is expected to be improved.

- Evaluation of risk impact can be calibrated, if there are few developers who have rich experience of risk. So, when making the risk list by questionnaire, it is better to collect risk experience and answers related to dread and risk factors for the calibration.

Our future work will involve collecting risk assessment of other kinds of risks using our questionnaire to evaluate the applicability of the results.

ACKNOWLEDGMENT

This research was partially supported by the Japan Ministry of Education, Science, Sports, and Culture [Grant-in-Aid for Scientific Research (C) (No. 16K00113)], and CAPES Brazilian Agency.

REFERENCES

[1] I. Amano, K. Kurisu, J. Nakatani, and K. Hanaki, "Effect of Provided Information and Recipient's Personality on Risk Perception of Drinking Water," Journal of Japan Society on Water Environment, vol.36, no.1 pp.11–22, 2013.

[2] H. Barki, S. Rivard, and J. Talbot, "Toward an Assessment of Software Development Risk," Journal of Management Information Systems, vol.10, no.2, pp.203–225, 1993.

[3] B. Boehm (Ed.), Software Risk Management, IEEE Press, Piscataway, NJ, USA, 1989.

[4] B. Boehm, "Software risk management: principles and practices," IEEE Software, vol.8, no.1, pp.32–41, 1991.

[5] B. Boehm and R. Ross, "Theory-W software project management principles and examples," IEEE Transactions on Software Engineering, vol.15, no.7, pp.902–916, 1989.

[6] R. Charette, Software Engineering Risk Analysis and Management, McGrawHill, Inc., New York, NY, USA, 1989.

[7] S. Du, M. Keil, L. Mathiassen, Y. Shen, and A. Tiwana, "The Role of Perceived Control, Attention-Shaping, and Expertise in IT Project Risk Assessment," In Proceedings of Annual Hawaii International Conference on System Sciences (HICSS), vol.8, 192c–192c, 2006.

[8] R. Gault, "A History of the Questionnaire Method of Research in Psychology," The Pedagogical Seminary, vol.14, no.3, pp.366–383, 1907.

[9] J. Hair, W. Black, B. Babin, and R. Anderson (Eds.), Multivariate Data Analysis (7th Ed.), Pearson, 2010.

[10] L. Hatcher, A Step-by-Step Approach to Using the SAS System for Factor Analysis and Structural Equation Modeling (1st ed.), SAS Publishing, 1994.

[11] S. Hirota, "Risk perceptions of death rates from hazards described as time scale versus as population scale," Japanese Journal of Social Psychology, vol.30, no.2, pp.121–131, 2014

[12] C. Howarth, "The relationship between objective risk, subjective risk and behaviour," Ergonomics, vol.31, no.4, pp.527–535, 1988.

[13] Project Management Institute, A Guide To The Project Management Body Of Knowledge (PMBOK Guides), Project Management Institute, 2004.

[14] R. Tatham J. Hair, R. Anderson and W. Black, Multivariate Data Analysis (4th Ed.): With Readings, Prentice-Hall, Inc., Upper Saddle River, NJ, USA, 1995.

[15] M. Jørgensen, The Value of Empirical Software Engineering Research, Presentation at Chalmers University,2010.

[16] Y. Kamei, A. Monden, S. Matsumoto, T. Kakimoto, and K. Matsumoto, "The Effects of Over and Under Sampling on Fault-prone Module Detection," In Proceedings of International Symposium on Empirical Software Engineering and Measurement (ESEM), IEEE Computer Society, Washington, DC, USA, pp.196–204, 2007.

[17] M. Keil, P. Cule, K. Lyytinen, and R. Schmidt, "A Framework for Identifying Software Project Risks," Communications of the ACM, vol.41, no.11, pp.76–83, 1998.

[18] M. Keil, L. Li, L. Mathiassen, and G. Zheng, "The Influence of Checklists and Roles on Software Practitioner Risk Perception and Decision-Making," In Proceedings of Annual Hawaii International Conference on System Sciences (HICSS), vol. 9, pp.229b–229b, 2006.

[19] M. Keil, L. Wallace, D. Turk, G. Dixon-Randall, and U. Nulden, "An Investigation of Risk Perception and Risk Propensity on the Decision to Continue a Software Development Project," Journal of Systems and Software, vol.53, no.2, pp.145–157, 2000.

[20] T. Kobayashi and T. Hoshino, "Propensity Score Adjustment for Internet Surveys of Voting Behavior," Japanese Journal of Electoral Studies, vol.27, no.2, pp.104–117, 2012.

[21] S. Leonard, "Classical and Modern Methods of Psychological Scale Construction," Social and Personality Psychology Compass, vol.2, no.1, pp. 414–433, 2007.

[22] K. Lyytinen, L. Mathiassen, and J. Ropponen, "Attention Shaping and Software Risk – A Categorical Analysis of Four Classical Risk Management Approaches," Info. Sys. Research, vol.9, no.3, pp.233–255, 1998.

[23] F. McFarlan, Software Risk Management, IEEE Press, Piscataway, NJ, USA, Chapter Portfolio Approach to Information Systems, 17–25, 1989.

[24] M. Norris and L. Lecavalier, "Evaluating the Use of Exploratory Factor Analysis in Developmental Disability Psychological Research," Journal of Autism and Developmental Disorders, vol.40, no.1, pp.8–20, 2010.

[25] S. Pilot, "What is fault tree analysis," Quality Progress, vol.35, no.3, pp.120–127, 2002.

[26] C. Piney, Risk identification: combining the tools to deliver the goods, Paper presented at PMI Global Congress 2003: Project Management Institute, 2003.

[27] R. Rabechini and M. Carvalho, "Understanding the Impact of Project Risk Management on Project Performance: An Empirical Study," Journal of Technology Management & Innovation, vol,8, pp.64–78, 2013.

[28] J. Ropponen and K. Lyytinen, "Can software risk management improve system development: an exploratory study," European Journal of Information Systems, vol.6, no.1, pp.41–50, 1997.

[29] J. Ropponen and K. Lyytinen, "Components of software development risk: how to address them? A project manager survey," IEEE Transactions on Software Engineering, vol.26, no.2, pp.98–112, 2000.

[30] M. Sapp, Basic Psychological Measurement, Research Designs, And Statistics Without Math, Charles C Thomas Pub, 2006.

[31] R. Schmidt, K. Lyytinen, M. Keil, and P. Cule, "Identifying Software Project Risks: An International Delphi Study," J. Manage. Inf. Syst., vol.17, no.4, pp.5–36, 2001.

[32] D. I. K. Sjoeberg, J. E. Hannay, O. Hansen, V. B. Kampenes, A. Karahasanovic, N. Liborg, and A. C. Rekdal, "A survey of controlled experiments in software engineering," IEEE Transactions on Software Engineering, vol.31, no.9, pp.733–753, 2005.

[33] P. Slovic, "Perception of Risk," Science, vol.236, no.4799, pp.280–285, 1987.

[34] J. Walker and P. Almond, Interpreting Statistical Findings: a guide for health professionals and students, Open University Press, 2010.

[35] S. Walters, Quality of Life Outcomes in Clinical Trials and Health-Care Evaluation: A Practical Guide to Analysis and Interpretation, Wiley, 2009.